**Internally Oxidized Nb₃Sn Strands with Fine Grain Size and High Critical Current Density**

*Xingchen Xu, Michael D. Sumption*, and Xuan Peng*

X. Xu, Prof. M. D. Sumption
Department of Materials Science and Engineering, the Ohio State University, Columbus, OH 43210, USA
E-mail: sumption.3@osu.edu
Dr. X. Peng
Hyper Tech Research Incorporated, 539 Industrial Mile Road, Columbus, OH 43228, USA

Nb₃Sn superconducting strands are the most practical conductors to generate high magnetic fields (12-16 T), and thus have significant applications in nuclear magnetic resonance (NMR), and great potential for fusion reactors and particle accelerator magnets. High critical current density ($J_c$) is a key parameter for such applications. Significant efforts towards optimization of various factors led to an 80% improvement in $J_c$ from the early 1990s to 2003, when the 4.2 K, 12 T non-matrix $J_c$ reached 3000 A/mm² (corresponding to 5000 A/mm² in Nb₃Sn layer $J_c$).[1,2] However, further efforts over the past decade have failed to bring about further increase beyond this level,[3,4] leading some researchers to conclude that the $J_c$ of conventional Nb₃Sn strands had reached its maximum. Here, however, by applying an internal oxidation method, we reduce the grain size by a factor of three and nearly double the 12 T $J_c$. In this method, a Nb₃Sn strand is fabricated with Nb-Zr alloy as starting material; with oxygen supplied properly via an oxide powder, the Zr atoms in the Nb-Zr alloy are internally oxidized, forming fine intra-granular and inter-granular ZrO₂ particles in Nb₃Sn layer, which effectively refine Nb₃Sn grain size. At a reaction temperature of 625 °C, grain size down to 20-50 nm (36 nm on average) has been achieved. For this sample the 4.2 K, 12 T Nb₃Sn layer $J_c$ reached 9600 A/mm².

For $Nb_3Sn$, improvement in critical current density $J_c$ can be realized by refining grain size.[5,6] Present-day $Nb_3Sn$ strands have average grain sizes typically of 100-200 nm.[6] Previous work on $Nb_3Sn$ thin films fabricated by electron-beam co-evaporation deposition showed that by refining the grain size down to 15-30 nm, the high-field $J_c$ of $Nb_3Sn$ will be markedly enhanced.[7] At present the only practical method to refine grain size in $Nb_3Sn$ strands is to decrease the reaction temperature (a heat treatment above 600 °C is required for $Nb_3Sn$ phase formation in a $Nb_3Sn$ strand, which is composed of Nb or Nb-Ta, Sn, and Cu precursors in the green state). However, even reducing the temperature down to 615 °C can only refine the grain size to about 90 nm,[6] far from the goal of sub-30 nm. In this communication we show that an internal oxidation method can be the key to realize this goal. For a solid solution with the solute less noble than the solvent, oxygen can be supplied such that the oxygen partial pressure is high enough to oxidize the solute but not high enough to oxidize the solvent, in which case the solute will be selectively oxidized (internally oxidized).[8] The oxide product of the solute may precipitate out in the matrix (solvent) in the form of fine particles, which may be used for dispersion strengthening or grain refinement.[8,9] To apply this method to a $Nb_3Sn$ strand, commercially available Nb-1wt.%Zr alloy (in place of the usual Nb or Nb-Ta alloy) could be used because Zr, according to the Ellingham diagram,[10] has much stronger affinity to oxygen than Nb does, making internal oxidation of the alloy possible.[9,11] Oxygen supply, on the other hand, needs special attention. Oxygen cannot be delivered to a $Nb_3Sn$ strand externally, because the outer Cu sheath will block oxygen transport to the Nb-Zr alloy (this is clear from a kinetics calculation, and is verified by our experiments). Pre-dissolving oxygen in the Nb-1Zr alloy is not feasible either, because the precursor Nb-Sn composite is processed down from a large billet to a small diameter (typically tens of μm) filament, and the relatively high hardness of oxygen-containing Nb-Zr makes processing difficult.[11] In this work, oxide powder is added into the composite to supply oxygen within the strand.[11,12] Selection of the oxide (here denoted $MO_n$) is critical.

On one hand, M should be more noble than Nb, so that during heat treatment the Nb-1Zr can reduce $MO_n$ and take up the oxygen. On the other hand, M should not be too noble, because the oxygen supply rate by such oxides could be high enough to oxidize both Nb and Zr non-selectively, in which case surface oxidation occurs instead of internal oxidation – this has been observed in our experiments. We have determined from our studies that $SnO_2$, ZnO, and $Nb_2O_5$ are workable choices. To prevent Cu (the Cu itself is needed to promote $Nb_3Sn$ formation)[13] blocking the oxygen transfer path from the oxide to the Nb-1Zr,[12] the oxide powder is located adjacent to the Nb-1Zr alloy, and to guarantee the flowability during wire drawing, nano-metric powder was used. **Figure 1 (a) and (b)** show back-scatter electron (BSE)/scanning electron microscopy (SEM) images of the polished surfaces of a strand fabricated with $SnO_2$ powder located between the Cu/Sn core and Nb-1Zr tube wall, in the green state and after a reaction at 650 °C for 400 hours, respectively. A similar strand which used $NbO_2$ powder instead of $SnO_2$ was also fabricated. Because $NbO_2$ supplies little oxygen (<0.3 at.%) to the Nb-1Zr alloy (note that $Nb_2O_5$ is the only Nb-O compound that can supply considerable amount of oxygen to Nb alloy), while $SnO_2$ can supply >2 at.% O to fully oxidize the Zr atoms,[12] the strand with $NbO_2$ powder can be used a control. **Figure 1 (c) and (d)** show high-magnification secondary electron (SE)/SEM images of the fracture surfaces of the strands with $NbO_2$ and $SnO_2$, both reacted at 625 °C for 800 hours. The grain size distribution of the strand with $SnO_2$ (calculated from Figure 1d) is mainly within 20-50 nm, with an average of 36 nm, while that of the strand with $NbO_2$ (Figure 1c) is mainly within 30-160 nm, averaging out at 81 nm, which is slightly smaller than conventional $Nb_3Sn$ strands reacted at 625 °C (100-110 nm),[14] perhaps because the small amount of oxygen released by the $NbO_2$ contributed to the refinement of grain size. The above data show that internal oxidation in the strand with $SnO_2$ not only markedly refined the grain size, but also made the span of grain size distribution relatively smaller.

The grain size refinement is due to formation of $ZrO_2$ particles: as oxygen diffuses into the Nb-Zr alloy, O atoms are trapped by Zr atoms, forming Zr-2O clusters,[9] which subsequently precipitate out in the form of fine $ZrO_2$ particles as the Nb reacts with Sn to form $Nb_3Sn$. A transmission electron microscopy (TEM) image displaying both intra-granular and inter-granular $ZrO_2$ particles in an internally oxidized sample is shown in **Figure 2**. The particle size is about 5-20 nm, perhaps even smaller in samples reacted at lower temperatures. There are two possible mechanisms for grain refinement by $ZrO_2$ particles. First, these $ZrO_2$ particles may serve as nucleation centers for the grains during $Nb_3Sn$ layer growth, which would increase the number of grains and decrease their average size. Second, fine $ZrO_2$ particles may pin the $Nb_3Sn$ grain boundaries, preventing their migration and associated grain coarsening. Since grain coarsening is a process in which larger grains grow at the expense of their neighboring smaller grains, the grain-boundary-pinning effect of the $ZrO_2$ particles may also explain the more uniform distribution of grain size in such samples. However, we also notice that in the internally oxidized samples grain coarsening may occur for long reaction times. SEM images of the strand with $SnO_2$ reacted at 650 °C for 150 h and 400 h are shown in **Figure 1 (e) and (f)**, respectively. After a 150-hour reaction, all the grains were very fine; however, when the reaction time was prolonged to 400 hours, some grains (denoted "unrefined" in Figure 1f) grew to sizes similar to those of conventional $Nb_3Sn$ strands. This perhaps indicates that the refined grains associated with $ZrO_2$ pinning are in fact in a metastable state. This may be due to the coarsening of the $ZrO_2$ particles themselves for long reaction times, and a corresponding local coarsening of the grain size. More TEM studies are needed to verify this. Fortunately, practical $Nb_3Sn$ filaments are much smaller and thus do not require such long reaction times.

Magnetic moment vs field (*m-B*) loops were then measured for these samples at 4.2 K, and the $Nb_3Sn$ layer $J_c$s were calculated based on the expression for a hollow cylinder in a perpendicular field, $J_c=3\Delta m/\{L*(d_o^3-d_i^3)\}$, where $\Delta m$ is the height of the *m-B* loop, *L* is the sample length, $d_o$ and $d_i$ are the outer and inner diameters of the current-carrying Nb$_3$Sn layer, respectively.[6] The $d_o$ was measured from the whole-strand BSE image (e.g., Figure 1b); but the inner boundary cannot be easily discerned due to the presence of a layer of ill-connected Nb$_3$Sn coarse grains, which are transformed from an intermediate Nb-Sn phase and do not carry transport current.[13] Thus, we measured Nb$_3$Sn layer thickness $t$ from the SE images of fracture surfaces (e.g., Figure 1f), and extracted $d_i$, since $d_i=d_o-2t$. To minimize error, several images were taken from various positions of the cross section, and on each image at least 6 line measurements were performed; $t$ was calculated by averaging these measurements, with the error estimated to be smaller than 5%. The calculated layer $J_c$s are shown in **Figure 3**, along with the $F_p$-$B$ curves calculated from them ($F_p=J_c\times B$). The 4.2 K, 12 T layer $J_c$s of the strand with NbO$_2$ reacted at 625 °C for 800 h is 4400 A/mm$^2$, close to those of the present-day Nb$_3$Sn strands. On the other hand, the values of the strand with SnO$_2$ reacted at 650 °C for 400 h and at 625 °C for 800 h are 8500, and 9600 A/mm$^2$, respectively, with the latter one almost doubling that of the best present-day Nb$_3$Sn strands. We also measured the transport $J_c$ using the four-point method in liquid helium, but because the Nb$_3$Sn layer is too large ($d_o$ >170 μm) the measurement quenched before the critical current was reached.[15] The measured 10 T layer current density at quench for the strand with SnO$_2$ (650 °C / 400 h) was 10,600 A/mm$^2$.

The $F_{p,max}$ (the peak values of the $F_p$-$B$ curves shown in Figure 3b) vs the reciprocal of grain size, $1/d$, of the internally oxidized samples are plotted in **Figure 4 (a)**, along with data for conventional Nb$_3$Sn.[6,16] It is interesting to note that $F_{p,max}$ increases linearly with $1/d$ when grain size is large, but the curve bends over as the grain size goes below ≈100 nm. Subsequently, the $F_p$-$B$ curves in Figure 3 (b) were fitted to a universal scaling law to obtain the $B_{irr}$ values ($B_{irr}$ is the magnetic field at which $J_c$ drops to zero).[12,17] The $B_{irr}$ value for the strand with NbO$_2$ (625 °C/800 h) is 20.9 T, while those of the strand with SnO$_2$ are 23 T (for 650 °C/400 h) and 19.4 T (for 625 °C/800 h). Using these, normalized $F_p$-$B$ curves were generated and are shown in **Figure 4 (b)**, from which we see that the strand with $NbO_2$ peaks at $\approx 0.2B_{irr}$, following the common characteristic of conventional $Nb_3Sn$ strands.[18] For the strand with $SnO_2$ (650 °C/400h), the average grain size of which is 43 nm, we see a peak at $0.26B_{irr}$, indicating a slight shift to higher field. The shift for the strand with $SnO_2$ (625 °C/800 h) is, however, quite clear, at $0.34B_{irr}$. Herein, we conclude that when grain size is above $\approx$50 nm, a decrease of the grain size only shifts the $F_p$-$B$ curve upward (while the peak remains at $0.2B_{irr}$). However, as grain size goes below a threshold of 30-50 nm, not only does $F_{p,max}$ increase, but also the $F_p$-$B$ curve peak shifts to higher fields. Some possible explanations for such an effect were discussed elsewhere.[19,20] Using the above fitting, we estimate the 4.2 K, 15 T layer $J_c$ of the strand with $SnO_2$ (650 °C / 400 h) to be 3900 A/mm$^2$, about 50% higher than that of the best present-day $Nb_3Sn$ strands.[4]

In addition, we find that the internal oxidation also influences the chemical reaction rate. Looking to the SEM images of the internally oxidized and control samples reacted at 750 °C and 850 °C shown in **Figure 5**, we see that the internally oxidized sample reacted at 750 °C had a smaller $Nb_3Sn$ layer thickness than the control sample (similar phenomena were also observed at 650 °C and 625 °C), while at 850 °C the opposite was observed, suggesting that internal oxidation decreased the layer growth rate at low reaction temperatures, but enhanced it at high temperatures. One possible explanation for this phenomenon is as follows. The growth of the $Nb_3Sn$ layer is a process of Sn diffusion through the $Nb_3Sn$ layer and reaction of Sn with Nb at $Nb_3Sn$/Nb interface. Since internal oxidation refines the $Nb_3Sn$ grain size, it should enhance the diffusion rate, given that Sn mainly diffuses along $Nb_3Sn$ grain boundaries.[21] On the other hand, we expect the reaction rate to be reduced due to the process associated with the re-organization of the dispersed Zr-O clusters into $ZrO_2$ particles during the Nb-Sn reaction (more TEM studies are required to confirm this speculation). Previous experiments suggested that the $Nb_3Sn$ layer growth shifts from reaction-rate limited to diffusion-rate limited as the reaction temperature is increased.[22] Therefore, internal oxidation should decrease the layer growth rate at low reaction temperatures but enhance it at high temperatures. In addition, it is expected that the dispersed fine $ZrO_2$ particles also influence other properties (e.g., mechanical properties) of these strands.

The internal-oxidation-driven improvements in $Nb_3Sn$ properties demonstrated above are compelling. Nevertheless, performance can still be pushed much higher. First, grain size can be further reduced with the use of even lower reaction temperatures (e.g., 605 °C, see [24]), or higher Zr content (e.g., 1.5%) to produce a greater density of $ZrO_2$ particles. Second, we can explore the addition of Ta or Ti dopant, which is well known to increase $B_{irr}$ in $Nb_3Sn$.[25] In this way we anticipate that we can attain the goal of refining grain size down to 15-30 nm while keeping $B_{irr}$ above 22 T.[24] Based on this, it is estimated that the 20 T whole-strand $J_c$ might surpass that of the recently-developed, record-$J_c$ Bi:2212 strands reacted at high over-pressures.[26]

In addition to the structure we show in Figure 1, this internal oxidation method can be easily transferred to other $Nb_3Sn$ strand architectures. For example, the oxide powder can be blended with other Sn-source powders in the core of powder-in-tube strands. Strands can also be fabricated by restacking Cu-sheathed Nb alloy rods (as in the rod-restack process) instead of using a Nb alloy tube:[1,23] in such a structure, oxygen can be supplied by distributing CuO powder-containing Cu tubes among Nb-1Zr rods, a structure similar to the one in which Nb-47Ti was introduced.[23] We have demonstrated the feasibility of supplying oxygen to Nb-1Zr through thin-layer Cu using CuO powder.

In summary, we show that for $Nb_3Sn$ strands we can reduce the grain size by a factor of three (down to 36 nm), shift the $F_p$-B curve peak to higher fields (0.34$B_{irr}$), and nearly double the 12 T $J_c$ (up to 9600 A/mm$^2$ at 4.2 K) by using an internal oxidation method, in which Zr atoms in Nb-Zr alloy are internally oxidized to form $ZrO_2$ particles. Further reduction of grain size and inclusion of ternary alloying would increase this further still, perhaps allowing $Nb_3Sn$ to compete with high-temperature superconductors at magnetic fields up to 20 T.

**Experimental Section**

For all the strands used in this work straight segments of about 25 cm in length were reacted in a furnace under flowing Ar. Samples for characterizations were cut from the centers of the reacted strands. A Sirion field emission SEM was used to obtain BSE images on sample surfaces polished to 0.05 μm and SE images on fractured surfaces, the latter used to determine $Nb_3Sn$ grain size. Specimens for TEM studies were prepared via focused ion beam (FIB) using an FEI Helios NanoLab 600. TEM images were taken on an FEI/Philips CM-200T TEM. Measurements of m-$B$ loops were performed in perpendicular magnetic fields with a ramp rate of 13 mT/s using the vibrating sample magnetometer (VSM) function of a Quantum Design Model 6000 "physical property measuring system" (PPMS).

**Supporting Information**
Supporting Information is available from the Wiley Online Library or from the author.

**Acknowledgements**
This work was funded by the US Department of Energy, Division of High Energy Physics, Grant No. DE-FG02-95ER40900, and a DOE Contract Numbers DE-SC0010312.

Received: ((will be filled in by the editorial staff))
Revised: ((will be filled in by the editorial staff))
Published online: ((will be filled in by the editorial staff))

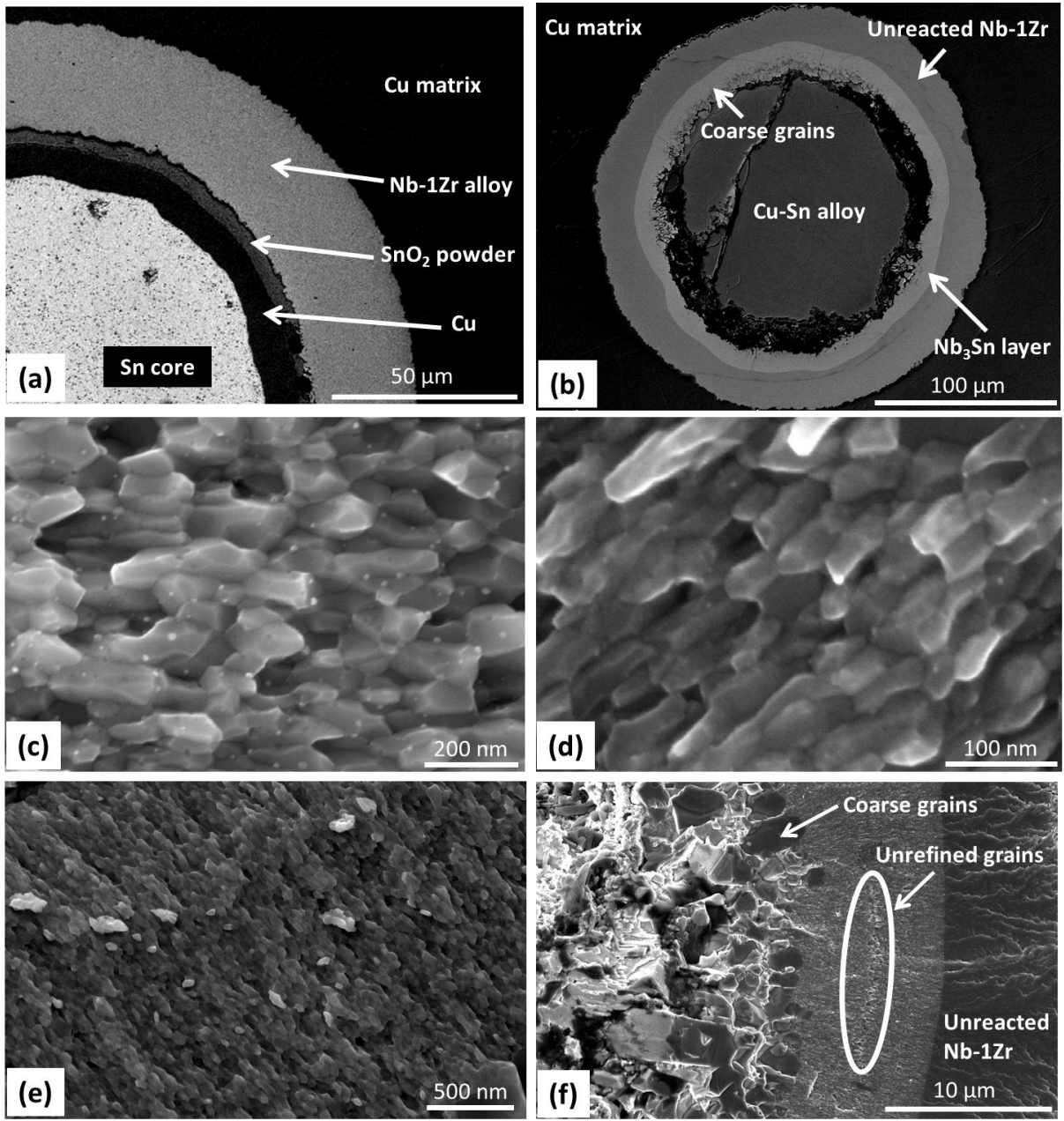

**Figure 1.** BSE/SEM images for the strand with SnO$_2$ powder, a) before reaction and b) after a reaction at 650 °C for 400 hours; SE images of fracture surfaces for c) the strand with NbO$_2$ reacted at 625 °C for 800 hours, and the strand with SnO$_2$ reacted at d) 625 °C for 800 hours, e) 650 °C for 150 hours, and f) 650 °C for 400 hours.

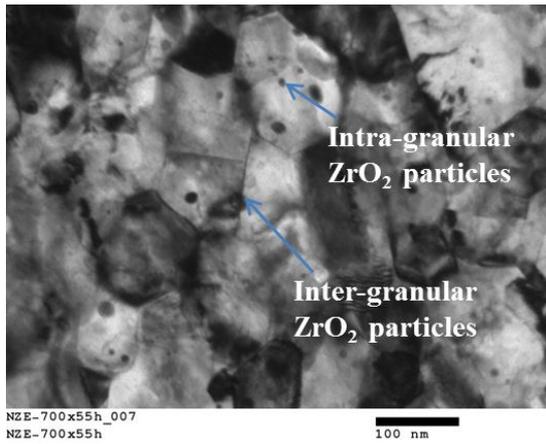

**Figure 2.** TEM image of an internally oxidized sample reacted at 700 °C for 55 hours, showing the intra-granular and inter-granular $ZrO_2$ particles.

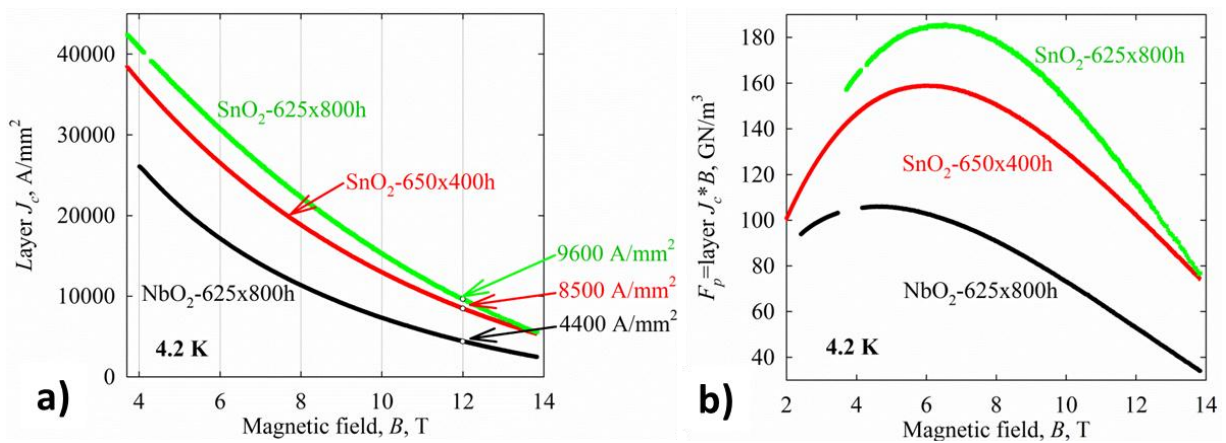

**Figure 3.** a) Layer $J_c$-$B$ curves and b) $F_p$-$B$ curves (4.2 K) for the strands with $NbO_2$ and $SnO_2$ given the specified reactions.

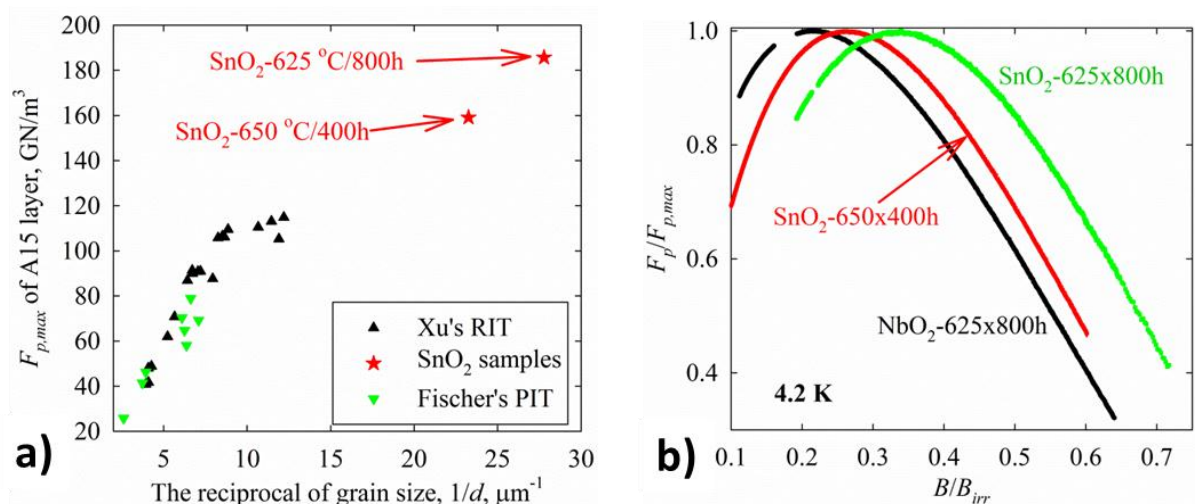

**Figure 4.** a) $F_{p,max}$ vs reciprocal grain size for the internally oxidized samples and some conventional $Nb_3Sn$,[6,16] and b) normalized $F_p$-$B$ curves (4.2 K) for the strands with $NbO_2$ and $SnO_2$ given the specified reactions.

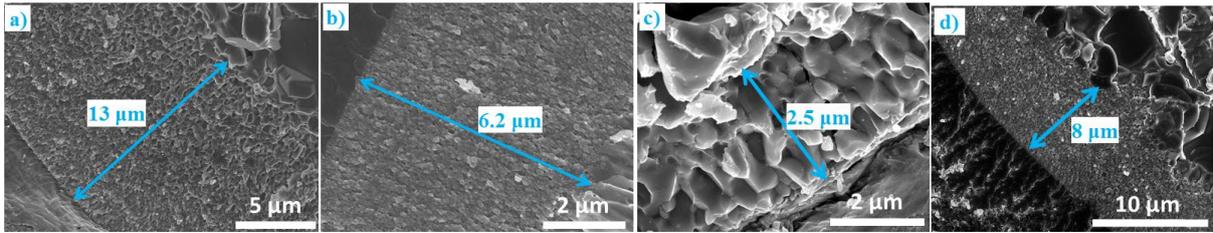

**Figure 5.** The a) control sample and b) internally oxidized sample, reacted at 750 °C for 8 hours, as well as the c) control sample and d) internally oxidized sample, reacted at 850 °C for 40 minutes.